\documentstyle[aas2pp4,flushrt]{article}

\newcommand\psr{PSR~J1814$-$1744}
\newcommand\axp{1E~2259$+$586}
\newcommand\asca{{\it ASCA}}
\newcommand\rosat{{\it ROSAT}}

\newcommand\LX{ergs~s$^{-1}$}
\newcommand\FX{ergs~s$^{-1}$~cm$^{-2}$}

\newcommand{\degrees}{\mbox{$^{\circ}$}}

\begin{document}

\title{X-ray observations of the high magnetic field radio pulsar PSR 
J1814$-$1744}

\author{M. J. Pivovaroff, V. M. Kaspi\altaffilmark{1}}
\affil{Department of Physics and Center for Space Research, Massachusetts Institute of Technology, Cambridge, MA 02139}

\authoremail{mjp@space.mit.edu, vicky@space.mit.edu}

\altaffiltext{1}{Alfred P. Sloan Research Fellow}

\and

\author{F. Camilo}
\affil{Columbia Astrophysics Laboratory, Columbia University, New York, NY 10027}

\authoremail{fernando@astro.columbia.edu}

\begin{abstract}
PSR~J1814$-$1744 is a 4~s radio pulsar with surface dipole
magnetic field strength $5.5 \times 10^{13}$~G, inferred assuming
simple magnetic dipole braking.  This pulsar's spin
parameters are very similar to those of anomalous X-ray pulsars (AXPs),
suggesting that this may be a transition object between the radio
pulsar and AXP population, if AXPs are isolated, high magnetic field
neutron stars as has recently been hypothesized.  We present archival X-ray
observations of \psr\ made with \rosat\ and \asca.  X-ray emission is
not detected from the position of the radio pulsar.  The derived upper
flux limit implies an X-ray luminosity significantly smaller than those
of all known AXPs.  This conclusion is insensitive to the possibility
that X-ray emission from \psr\ is beamed or that it undergoes modest
variability.  When interpreted in the context of the
magnetar mechanism, these results argue that X-ray emission from AXPs
must depend on more than merely the inferred surface magnetic field
strength.  This suggests distinct evolutionary paths
for radio pulsars and AXP, despite their proximity in period--period 
derivative phase space.

\end{abstract}

\section{Introduction}
Recently, \psr, an isolated radio pulsar with period $P = 4$~s and large period
derivative ($\dot{P} =  7.4 \times 10^{-13}$) was discovered (Camilo et
al.  2000, see also Manchester et al. 2000 and Kaspi et al. 2000) \nocite{ckl+00} 
\nocite{mlc+00} \nocite{kcl+00} in an ongoing survey of the Galactic
Plane for radio pulsars using the 64-m Parkes telescope (Lyne et al. 2000).
\nocite{lcm+00} The pulsar's surface magnetic field strength $B$, inferred 
under the assumption of a dipole rotating 
{\it in vacuo} (\cite{mt77}) from
\begin{equation}
\label{eq:1}
 B = 3.2 \times 10^{19} (P \dot{P})^{1/2}~{\rm G}
\end{equation}
is $5.5 \times 10^{13}$~G.  

This pulsar's properties are particularly interesting because they are
similar to those of anomalous X-ray pulsars (AXPs).  AXPs have spin
periods  $P = 6-12$~s, and spin down regularly (see, e.g., \cite{ms95}
, \cite{gv98}, and \cite{kcs99}) with period derivatives 
$10^{-12} < \dot{P} < 10^{-11}$.  The X-ray luminosities of AXPs are typically
several orders of magnitude larger than their spin-down luminosities \(
\dot{E} \equiv 4 \pi^{2} I \dot{P}/P^{3}\) (where $I$ is the pulsar's
moment of inertia, assumed to be $10^{45}$~g~cm$^{2}$) (see, e.g.,
Oosterbroek et al. 1998 and references therein).  Strong observational
evidence (see, e.g., \cite{mis98}) precludes accretion from a binary
companion as the origin of the observed X-rays.  Instead, the leading
hypothesis to explain AXP properties is that they are isolated neutron
stars with ultra-high magnetic fields, so-called ``magnetars'' (Duncan
\& Thompson 1992).  \nocite{dt92a} In this model, the X-ray emission is
powered either by decay of the large magnetic field (\cite{td96a}) or neutron
star cooling enhanced by the presence of the strong field 
(Heyl \& Hernquist 1997a).  \nocite{hh97} Assuming magnetic dipole 
braking, AXPs have inferred
surface dipole magnetic field strengths $B = (0.6 - 8) \times 10^{14}$~G.

While \psr\ is a radio pulsar and hence an isolated neutron star, its
spin parameters and hence inferred magnetic field are extreme in the
pulsar population:  its magnetic field is nearly three times larger
than that of PSR~B0154$+$61 (Arzoumanian et al. 1994), \nocite{antt94}
the pulsar with the previously known highest field
strength\footnote{The survey that discovered \psr\ also discovered
PSR~J1119$-$6127, a radio pulsar with field $B = 4.1 \times
10^{13}$~G (Camilo et al. 2000).}.  Given that the spin parameters of \psr\ are more typical
of AXPs than radio pulsars, the possibility that this is a transition
object between these two neutron star populations must be entertained.
In particular, under the magnetar hypothesis, the mechanism responsible
for the production of X-rays in AXPs should be present in \psr\ if the
inferred magnetic field is indeed the primary characteristic relevant
to the observed magnetar properties.

The similarity in spin parameters between PSR J1814$-$1744 and the AXPs is readily
seen in a ``$P$$-$$\dot{P}$'' plot.  Figure~1 shows $\dot{P}$ versus
$P$ for the radio pulsar population (small dots), with
\psr\ indicated.  The figure also shows the candidate magnetar
population, consisting of five AXPs  and two soft gamma repeaters
(SGRs).  Immediately noticeable is the proximity of \psr\ to the
cluster of AXPs and SGRs at the upper right corner.  It is especially
close to \axp, also indicated (\cite{kcs99} and references therein).  
Table~1 compares the properties of these two neutron stars.

\begin{figure}
\plotone{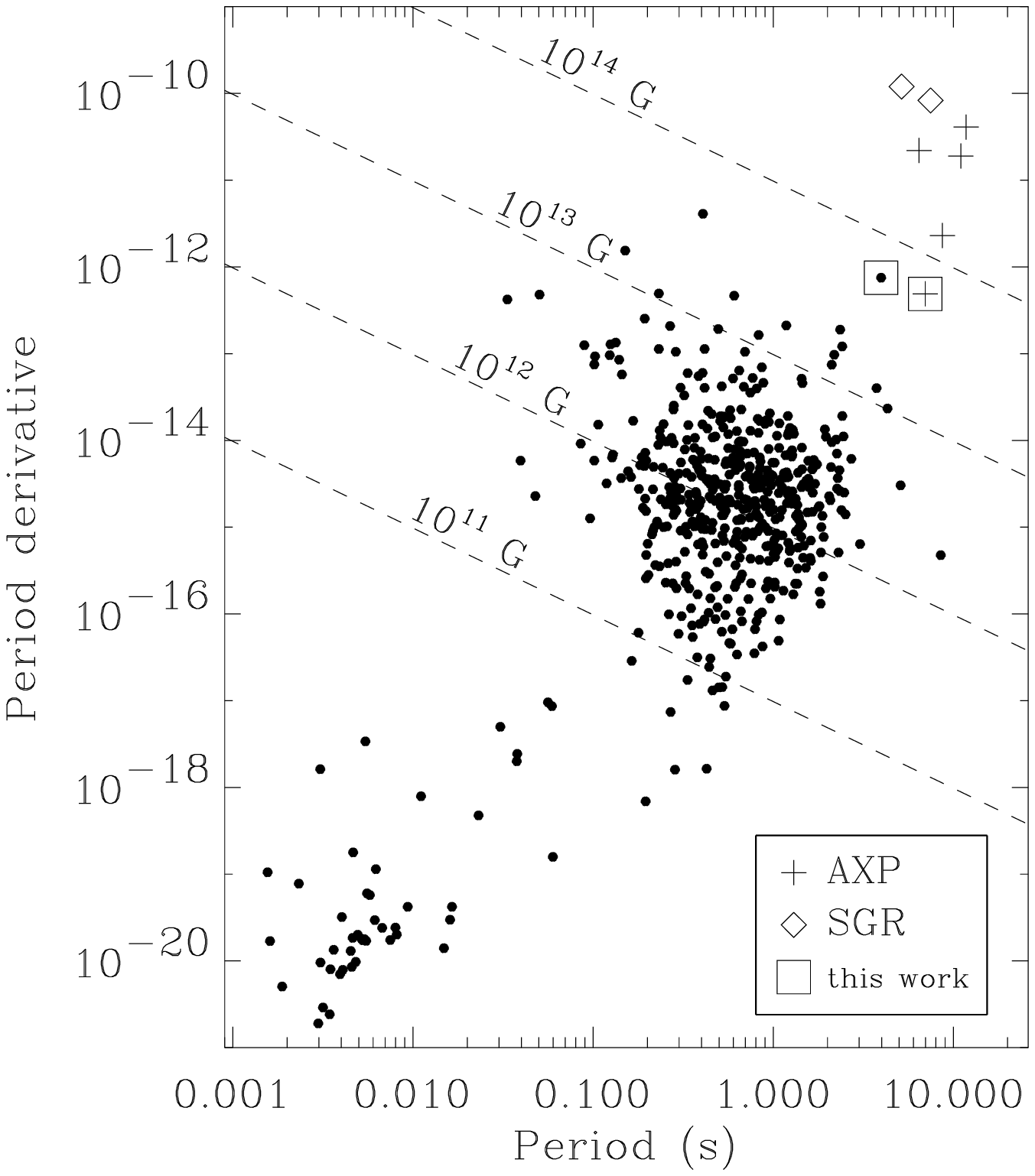}
\caption{$P-\dot{P}$ diagram.  The small dots are the known
radio pulsars.  The crosses are the AXPs and the diamonds are the SGRs.
Lines of constant magnetic field, derived from Equation~1, are shown by the
dashed lines.  Note the proximity of \psr\ (the boxed dot) to
\axp\ (the boxed cross).  Pulsar references--Taylor et al. 1995; 
Camilo et al. 2000; Young, Manchester \& Johnston 1999. 
SGR references--\cite{kds+98}; \cite{ksh+99}.  AXP references--see citations in the main text.}  
\label{fig1}
\end{figure}

\placefigure{fig1}

\placetable{tab1}

Here, we present an analysis of archival \rosat\ and \asca\ X-ray
observations that serendipitously include the location of \psr\ within
their respective fields of view (FOV).  Emission from the pulsar
position was not detected with either telescope.  We note that X-rays
from other mechanisms (e.g., magnetospheric emission) are not expected
to be observable from \psr.  Give its spin-down luminosity of $5 \times
10^{32}$~\LX\ and its probably distance of 10~kpc (estimated from the
distance/dispersion measure relation of Taylor \& Cordes [1993]), its
spin-down flux ($\dot{E}/4 \pi d^{2}$) implies that any
rotation-powered high-energy emission should be too faint to detect
(Seward \& Wang 1988; Becker \& Tr\"{u}mper 1997). \nocite{bt97}
\nocite{sw88} Count rate limits from both observations are used to
derive an upper limit on the X-ray luminosity from \psr\ that is
significantly smaller than those measured for the known AXPs.  We then 
discuss the implications of the non-detection for current magnetar models.

\section{Archival Data Analysis}
\label{sec:analysis}
\subsection{{\boldmath $ROSAT$}}

\label{subsec:rosat}
A field containing \psr\  was observed 
with the Position Sensitive Proportional Counter (PSPC) instrument
aboard \rosat\ (Tr\"{u}mper 1983) \nocite{tru83}
during 1992~April~2--8, as part of a study of the Galactic
X-ray background (sequence RP900196N00).  The radio timing position of
\psr\ (\cite{ckl+00}), 
$\alpha$ (J2000) = $18^{\rm h}~14^{\rm m}~43.^{\rm s}0(2)$,
$\delta$ (J2000) = $-17\degrees~44^{\prime}~47^{\prime\prime}(23)$, 
is located $34^{\prime}$ from the optical axis.  The total live time
was 7.7~ks.  A $0.1-2.4$~keV broad-band flat-fielded
image was produced using an exposure map that accounts for vignetting
and instrumental structure.  The exposure map is a standard data
product generated by the NASA-maintained HEASARC during its analysis of
all {\it ROSAT} observations.  No X-ray emission is present.  To
calculate an upper limit on the count rate, we compare the counts
collected in an aperture of radius $0.\! ^{\prime}$8 centered on the
radio position to those in a concentric annulus with radii $4^{\prime}$
and $12^{\prime}$.  The aperture radius represents the theoretical half
energy width for a point source located 34$^{\prime}$ off-axis,
calculated using the {\tt FTOOL PCRPSF~v.2.0.7}.

We define the signal-to-noise ratio \( S/N \equiv S/\sigma_{s} \), with signal
\( S \equiv N_{s} - {\it b} \) and variance $\sigma_{s}^{2}$.
$N_{s}$ is the number of counts in the circular aperture, 
\( {\it b} \equiv \beta N_{b} \), where 
$N_{b}$ is the number of counts in the 
annulus and $\beta$ is the ratio of exposure
areas of the aperture and annulus.  This ratio 
accounts for the geometric size of the source aperture and background annulus 
as well as the effective exposure time of each sky element comprising the 
two regions.  Using standard error propagation, 
\( S/N = S/\sqrt{S + b(1+\beta)} \).
Requiring $S/N > 3$$\sigma$ yields a count rate upper limit in
the $0.1-2.4$~keV band of $< 3.2 \times 10^{-3}$~cps.  

\subsection{{\boldmath $ASCA$}}
\label{subsec:asca}
A field containing \psr\ was observed with 
\asca\ (Tanaka, Inoue, \& Holt 1994) \nocite{tih94} 
on 1996~April~9, as part of a Galactic plane survey
(sequence 54005040).  \asca\ consists of four co-aligned telescopes,
each of which has its own focal plane detector: two Gas Imaging
Spectrometers (GIS-2 and GIS-3) and two Solid-state Imaging
Spectrometers (SIS-0 and SIS-1).  The pulsar position falls
13$^{\prime}$ from the ASCA optical axis.  This puts the source at the
edge of the SIS detectors, limiting the utility of these data.  We do
not consider them further, focusing instead on the GIS instruments
which have a much larger ($25^{\prime}$) FOV.  The effective exposure
time for the GIS is $2 \times 11.5$~ks.  The image obtained from
combining data from both GIS cameras has been corrected for pointing
offset\footnote{This offset arises due to a systematic error with the
\asca\ star tracker.  The correction was applied using the {\tt FTOOL
offsetcoord} and the look-up table available at {\tt
http://legacy.gsfc.nasa.gov/docs/asca/coord/updatecoord.html}.} and
exposure and rebinned with a $45^{\prime\prime} \times
45^{\prime\prime}$ box-car function.  To avoid the large instrument
background at the edge of the detectors, we restricted  the image to a
circular FOV with $20^{\prime}$ radius.  The image is dominated by
scattered flux from the nearby ($34^{\prime}$) bright low mass X-ray
binary (LMXB) GX~13$+$1 (Vrtilek et al. 1991).  Due to its
better mirror performance, \rosat\ does not suffer from
this contamination problem.  \nocite{vms+91} A fan-like pattern,
consistent with that from a bright point source located 34$^{\prime}$
from the optical axis (see, e.g., \cite{gen95}), and a shadow from a
mirror quadrant boundary fall where expected, given the satellite roll
angle and the alignment of GX~13$+$1 with the optical axis\footnote{The
interested reader is referred to Serlemitsos et al.  (1995)
\nocite{sjs+95} for a detailed discussion of the stray light properties
of \asca.}.  No X-ray emission is present from the location of the
radio pulsar.

An upper limit on flux from the pulsar is derived following the same
prescription outlined in $\S$\ref{subsec:rosat}, with an increase in
the radius of the source aperture to $4.^{\! \prime}5$ to accommodate
the wider PSF of \asca.  Fortuitously, \psr\ is situated in the middle
of the boundary shadow, largely shielding the location of the pulsar
from the scattered flux of GX~13$+$1.  However, the contamination of
nearly the entire GIS FOV from the LMXB complicates the choice of a
background region.  We considered using blank-sky data from the same
region of the detectors that encompass the source aperture, which would
account for the instrumental and cosmic X-ray backgrounds.  However,
the blank-sky background would not include contributions from the
diffuse Galactic plane emission, and most importantly, would not
include the scattered flux from GX~13$+$1.  Hence, we relied upon a
relatively contamination-free region from this observation to calculate
the background, selecting a circular region with $4^{\prime}$ radius,
located the same distance off-axis as \psr\ and containing roughly the
same amount of scattered light as the source aperture.  Requiring  $S/N
> 3$$\sigma$ yields a count rate  upper limit in the $2-10$~keV band
of  $< 5.6 \times 10^{-3}$~cps.  These calculations were performed on
the unbinned data, and while our approach should mitigate most of the
effects of the contamination from GX~13$+$1, we recognize it is
impossible to fully account for it.  The above upper limit represents a
conservative estimate of the maximum count rate from \psr.

\section{Discussion}
\label{sec:compare}
\subsection{The X-ray luminosity upper limit of PSR J1814$-$1744} 
The dependence of the magnetar hypothesis for AXP emission on
inferred magnetic field, coupled with the similarities in the spin
parameters of PSR J1814$-$1744  and \axp, suggests that PSR J1814$-$1744 
 may also share
similar X-ray properties.  Here, we use the spectral properties of
\axp\ as a template to convert the count rate upper limits from
\psr\ into an X-ray luminosity upper limit.  

The spectrum of \axp\ is
well-studied (Iwasawa, Koyama \& Halpern  1992; Corbet et al. 1995; Rho
\& Petre 1997; Parmar et al.  1998) \nocite{ikh92} \nocite{cso+95}
\nocite{rp97} \nocite{pof+98} and best described by a two-component
model, consisting of a black body and power law.  Rho \& Petre (1997)
combine data from \rosat, \asca, and {\it BBXRT} (Serlemitsos et al.
1992) \nocite{ser92} to determine best-fit parameters of $\Gamma = 4.0$
for the power law index and $kT = 0.43$~keV for the black body
temperature.  See $\S$3.1 for the distribution of the
spectral properties of all AXPs.  The unabsorbed flux is 
$F_{2-10~{\rm keV}} = 1.3 \times 10^{-11}$~\FX\ and 
$F_{0.1-2.4~{\rm keV}} = 4.1 \times 10^{-9}$~\FX.
The distance to \axp\ has been estimated to be $d = 3.6 - 5.6$~kpc from
the supernova surface brightness-distance ($\Sigma$-$D$) relationship
(\cite{gf80}; \cite{sth83}; \cite{hhc+84}).  While there is great
uncertainty in the $\Sigma$-$D$ relation (see, e.g., \cite{gre84} and
\cite{ber86}), distances to stars in nearby H{\sc ii} regions are of similar
value (Rho \& Petre 1997), resulting in a commonly quoted distance of
4~kpc.

In order to scale the spectral model of \axp\ for \psr, we must first
estimate the Galactic absorption (column density $N_H$) and distance to
the radio pulsar.  There are two coarse, yet independent, methods for
estimating $N_{H}$.  The Seward \& Wang (1988) approximation of 10 
neutral hydrogen atoms per free electron, combined with the pulsar
dispersion measure DM~= 834~pc~cm$^{-3}$ (Camilo et al. 2000) gives
$N_{H} =  2.6 \times 10^{22}$~cm$^{-2}$.  The {\tt FTOOL nh}, which
uses the H{\sc i} maps of Dickey \& Lockman (1990), \nocite{dl90} predicts
$N_{H} =  1.8 \times 10^{22}$~cm$^{-2}$.  The DM$-$$d$ relationship of
Taylor \& Cordes (1993) \nocite{tc93} predicts a distance of $\sim$10~kpc.  
At this distance, the unabsorbed X-ray flux of \axp\ would be reduced to
$F_{2-10~{\rm keV}} = 2.1 \times 10^{-12}$~\FX\ and 
$F_{0.1-2.4~{\rm keV}} = 6.6 \times 10^{-10}$~\FX.

Figure~2 shows the expected count rates in both the \rosat\ PSPC 
and the \asca\ GIS  as a function of $N_{H}$. The
rates were calculated by using {\tt XSPEC (v.10)} to fold the spectrum
of \axp, normalized to a distance of 10~kpc, through the appropriate
instrument response matrices.  Calculations at several values
of $N_{H}$, shown by the symbols, were used to interpolate the count
rate as a continuous function of $N_{H}$.  The dashed lines in each
plot represent the measured upper limits.  Even when $N_{H}$ is allowed
to exceed the maximum estimate, the predicted count rates are well
above the measured upper limits.  Assuming a reasonable compromise
value for the column density of $N_{H} = 2.2 \times 10^{22}$~cm$^{-2}$,
the expected \rosat\ count rate is a factor of 13 higher than measured,
and the expected \asca\ count rate is a factor of 6 higher than
measured.  Scaling the flux by these ratios, we find for \psr, 
\rosat\ gives 
$L_{0.1-2.4~{\rm keV}} < 6.3 \times 10^{35} (d/10~{\rm kpc})^{2}$~\LX\
and \asca\ gives 
$L_{2-10~{\rm keV}} < 4.3 \times 10^{33} (d/10~{\rm kpc})^{2}$~\LX.  
Extrapolating the {\it ROSAT}-derived upper limit to the
{\it ASCA} band gives $L_{2-10~{\rm keV}} < 2.0 \times 10^{33} 
(d/10~{\rm kpc})^{2}$~\LX, in good agreement with the {\it ASCA}-derived
limit.

\begin{figure} 
\plotone{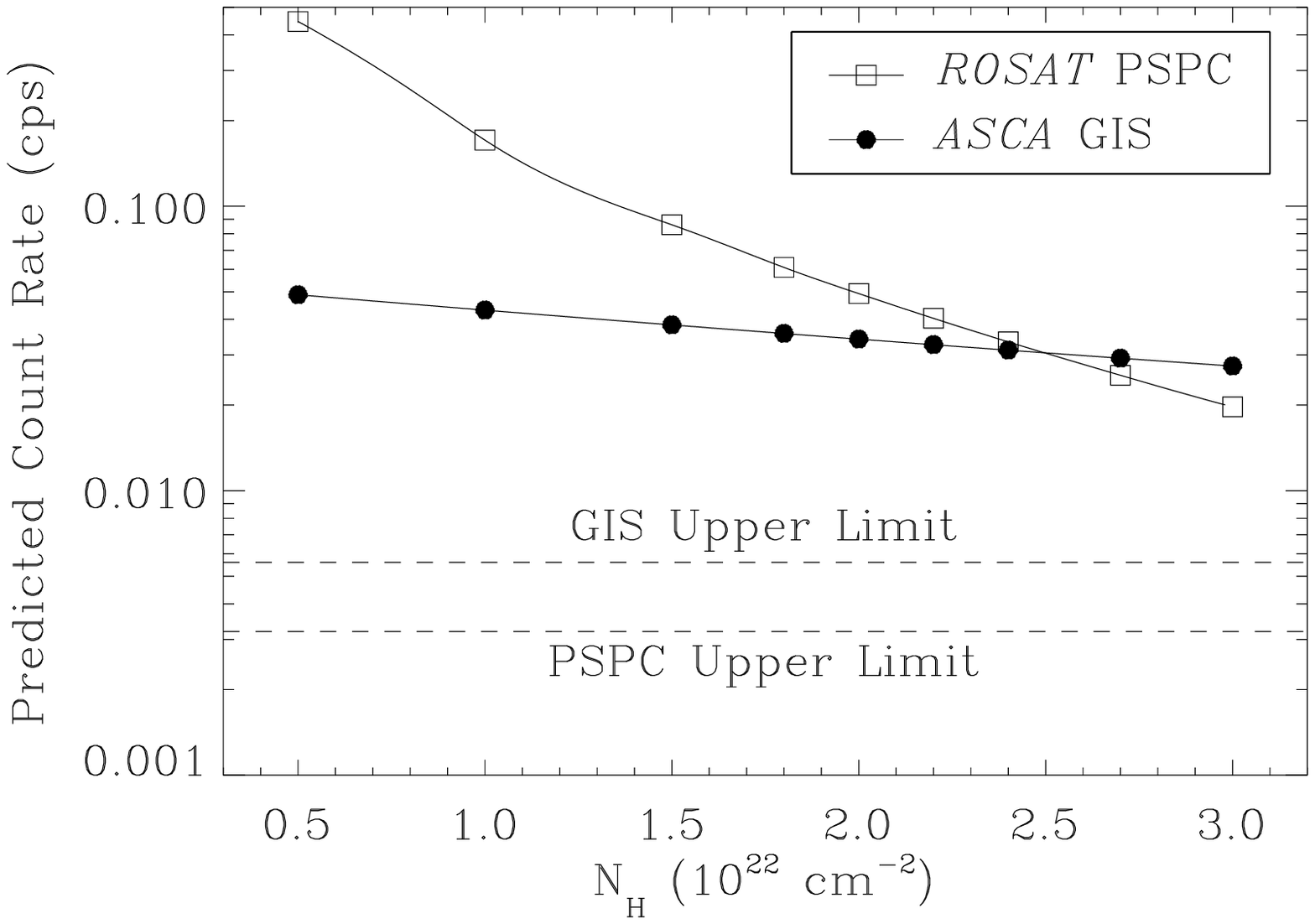}
\caption{The diagram shows the count rates expected if \psr\ had
the same spectrum and luminosity as \axp, as a function of $N_{H}$.
Calculations at several discrete values
of $N_{H}$, shown by the symbols, were used to interpolate the count
rate as a continuous function of $N_{H}$.  Distances of
10~kpc and 4~kpc were assumed for \psr\ and \axp, respectively.
The dashed lines represent the derived upper limit on count rate 
for the \rosat\ PSPC ($0.1-2.4$~keV) and the 
\asca\ GIS ($2-10$~keV).
Assuming a reasonable estimate for the column density of
$N_{H}= 2.2 \times 10^{22}$~cm$^{-2}$,
the expected count rate exceeds the upper limit by a factor
of 13 (PSPC) and 6 (GIS).}
\label{fig2}
\end{figure}

\placefigure{fig2}

Next, we comment on the use of the spectral properties of \axp\ for
calculating our luminosity upper limit for \psr.  The spectral
properties determined for AXPs are very similar, with power law photon
indices in the range $\Gamma = 2.5-4$ and black body temperatures in
the span $kT = 0.39 - 0.71$~keV (1E~2259$+$586: Rho \& Petre 1997;
1E~1048.1$-$5937: Oosterbroek et al 1998; 1E~1841$-$045: Vasisht \&
Gotthelf 1997; \\
RX~J170849.0$-$400910: Sugizaki et al 1997; 4U~0142$+$61: White et al
1996).  \nocite{wae+96} \nocite{vg97} \nocite{snt+97}
Their X-ray luminosities ($0.5-10$~keV) differ significantly, however,
varying by more than two orders of magnitude.  Calculations identical
to those described above show that if \psr\ had properties like the
other known AXPs, the expected count rates would all be greater than
those predicted from assuming the spectrum and luminosity of \axp.  We
note that only AXP 1E~1048.1$-$5937 (inferred surface magnetic field
strength $3.6 \times 10^{14}$~G, \cite{opmi98}) nominally yields a
lower expected count rate than the \axp\ model (though still higher than
our upper limit), although the
particularly large uncertainty in the distance of the former makes its
true intrinsic luminosity difficult to know (Corbet \& Mihara 1997).
\nocite{cm97}

Thus, if the measured spectral parameters for any of the other AXPs
(with the possible exception of 1E~1048.1$-$5937) were used in
interpreting the count rate upper limit for \psr, the difference
between its X-ray luminosity upper limit and the luminosity expected
from AXP-like emission would only increase.

\subsection{Beaming and Source Variability}
One scenario that must be considered for the lack of emission
from \psr\ is that the magnetar mechanism is present and emitting X-rays
that are beamed away from our line of sight.  The small number of
confirmed AXPs prevents any detailed statistical discussion of beaming.
However, the pulsed fraction and pulse shape of the known AXPs can be
used to motivate at least a rough characterization.  For each AXP,
Table~2 lists the pulsed fraction {\it f}, the passband and the pulse shape.
Here, we follow Page (1995) and estimate the pulsed
fraction using \( f = 1/2 \times ({\rm N_{max}} - {\rm
N_{min}})/{\rm N_{mean}} \), where the subscript refers to the
maximum, minimum, and mean counts in the pulse profile.  
All AXPs have large unpulsed components and have pulse 
profiles with smooth or sinusoidal shapes, independent of the number of
X-ray pulses present.  
Another general trend is that, excepting 4U~0142$+$61, there
is no evidence for appreciable pulse shape evolution with energy.

\placetable{tab2}

These properties are consistent with the predictions of both magnetar
models:  that pulsed emission from AXPs is best interpreted as smoothly
modulated thermal emission from the surface of the neutron star
(\cite{td96a}; \cite{hh97}). See $\S$3.3 for additional discussion.
The pulsed high energy component ($E > 2$ keV), well fit by a power-law
model, raises the possibility that non-thermal magnetospheric processes
may also contribute. The lack of AXP pulse evolution with energy,
however, suggests that only a single X-ray emission mechanism is
present in AXPs.  This is in contrast to rotation-powered pulsars like
the Vela pulsar, which has a soft, sinusoidal thermal component with
low pulsed fraction (\cite{ofz93}) and a sharply peaked non-thermal
component that extends above 100~MeV (Strickman, Harding, \& de Jager
1999).  \nocite{shd99} Of course, AXPs and rotation-powered pulsars
could have very different pulsed properties, given the uncertainty in
the origin of high-energy magnetospheric emission or the effect that a
high magnetic field would have on this mechanism.  For example,
distinct thermal and non-thermal components could be present yet, for
reasons unexplained, are locked in phase, thus appearing to have a
common origin.

In either magnetar model, the X-ray emission comes from the hot neutron
star surface.  Thus, it seems improbable that \psr\ would be oriented
with respect to our line of sight as to make X-ray pulsations
undetectable, especially considering that gravitational lensing can
make much more than half of the neutron star surface visible (see,
e.g., Page 1995 and Heyl \& Hernquist 1998).  Note, however, any
magnetospheric emission would be too distant from the neutron star
surface to undergo gravitational lensing, hence could be beamed.

But even if \psr\ were aligned in such a way, considerable unpulsed
X-ray emission would be present, as is the case of all the known
members of the AXP population (see Table~2).  For example, the unpulsed
luminosity of 1E~1841$-$045 is $\sim$85\% of the total observed
luminosity (\cite{vg97}).  Even if half of the flux is present in
pulsations directed away from our line of sight, the expected count
rates from \psr\ are still well above the upper limits derived from the
archival data.  We conclude that, even in the unlikely event that the
X-ray emission is significantly beamed, \psr\ should have been
detectable as an X-ray point source under the magnetar hypothesis.

Finally, we discuss the ramifications of source variability in AXPs.
Torii et al. (1998) \nocite{tkk+98} report variations in flux greater
than a factor of ten in the AXP candidate AX~J1844.8$-$0258.  Recent
observations by Vasisht, Gotthelf, Torii, \& Gaensler (2000)
\nocite{vgtg00} confirm the extreme variability 
of this object.  If this source is firmly established as an AXP by an
accurate $\dot{P}$ measurement and \psr\ has similar characteristics,
there exists the possibility that both serendipitous observations of
\psr\ occurred during a low-state and that it is, in fact, an X-ray
source.  If instead \psr\ undergoes variability by a factor of $2-3$,
as witnessed in \axp\ (Corbet et al. 1995 and references therein)
\nocite{cso+95} and 1E~1048.1$-$5937 (Corbet \& Mihara 1997),
\nocite{cm97} X-ray emission still would have been detected.

\subsection{Implications for magnetar models}
\label{subsec:discuss}
X-ray emission from AXPs has been explained in the context of the
magnetar model by either magnetic field decay (Thompson \& Duncan 1996)
\nocite{td96a} or neutron star cooling (Heyl \& Hernquist 1997a). \nocite{hh97}
Duncan \& Thompson (1992) \nocite{dt92a} first posited the existence of
magnetars to explain a subset of gamma-ray bursts, including those from
the SGRs.  Later (see, e.g., Duncan \& Thompson 1994 and Thompson \&
Duncan 1995), \nocite{dt94} \nocite{td95} they discuss how the
1979~March~5 event from SGR~0526$-$66, which released
$\sim$$10^{44}$~ergs in $\sim$0.2~s (\cite{cdp+80} and \cite{ekl+80}),
 could have been driven by a large-scale instability and subsequent
readjustment of the neutron star's magnetic field.  For the time scale
relevant to the system ($\sim$10$^{4}$~yr), they argue that the instability
is best explained by ambipolar diffusion of the magnetic field out of
the stellar core (\cite{td96a}).  This process conducts energy from the
core to the surface to produce X-ray luminosities in the range $L_{x}
\sim 10^{35}-10^{36}$~\LX.  The strong dependence of the surface heat
flux on field strength concentrates flux in the polar regions.  The
modulation of this intensity gradient gives rise to the X-ray
pulsations.

For 1E~2259+586, this mechanism can explain the observed X-rays if the
neutron star's characteristic age, $P/2\dot{P} = 230$~kyr, is an 
overestimate, the true age being $\sim 10$~kyr, that inferred for the 
proposed associated supernova remnant CTB~109 (Wang et al.
1992).  \nocite{wql+92} If this model is correct, the difference of
only 10\% in the magnetic fields of 1E~2259+586 and \psr\ seems
unlikely to account for the large difference in their X-ray
luminosities, given the similarity of the X-ray emission from the known
AXPs, whose magnetic fields span a much larger range.  Therefore, if
the decay of magnetic fields leads to the observed emission from AXPs,
that no X-rays are seen from \psr\ argues that \axp\ is much younger,
even though the characteristic ages indicate otherwise, consistent with
the latter's association with SNR CTB~109.  This suggests that \axp\ 
has had a spin-down history inconsistent with simple dipole braking or 
that it was born with a long spin period.

Heyl \& Hernquist (1997a) have suggested a second model, based on
photon cooling, for the production of X-rays in AXPs.  \nocite{hh97}
They extend into the magnetar regime ($B=10^{14}-10^{16}~{\rm G}$) the
work of Shibanov \& Yakovlev (1996), \nocite{sy96} who have shown that
when the magnetic field of a neutron star exceeds 10$^{12}$~G,
enhancements in the conductivity along the field lines due to electron
energy quantization result in a net increase in heat flux (Heyl \&
Hernquist 1997b; \cite{hh98a}).  In sufficiently young stars (ages of
$\sim$1~kyr) this leads to a gradual increase in photon luminosity with
increasing magnetic field.  The composition of the neutron star
envelope has an even more dramatic effect; the luminosity transmitted
through a helium envelope is six times that transmitted through an iron
envelope, while that transmitted through a hydrogen envelope is ten
times that for iron.  They then argue that the total mass of an
insulating low-Z layer required for the observed AXP luminosity is
modest and easily attained from either fall-back following the
supernova explosion or accretion from the ISM, if the pulsar has a
large birth velocity or the density is high enough.  Heyl \& Hernquist
(1997b; 1998) consider how the flux varies across the surface of a
magnetar and find a strong angular dependence caused by the anisotropic
heat conduction in the outer layers of the neutron star.  The
temperature gradient, coupled with the limb darkening expected from
magnetized atmospheres (see Heyl \& Hernquist 1997b and references
therein), \nocite{hh97a} results in a modulation of the thermal flux.
The observed pulsed fraction may be reduced by gravitational lensing,
which makes more than half of the neutron star surface visible at a
given instance (see, e.g. Page 1995 and Heyl \& Hernquist 1998)
\nocite{pag95} \nocite{hh98a} and effectively smooths the flux gradient
discussed above in $\S$3.1.

This model accounts for both the intensity of the X-ray emission and
the pulsed fractions from 1E~1841$-$045 and 1E~2259$+$586 (Heyl \&
Hernquist 1997a), \nocite{hh97} assuming the pulsar ages are those of
their associated supernova remnant, Kes~73 and CTB~109, respectively.
The absence of X-rays from \psr\ again argues that this neutron star
must be much older than 1E~2259+586.  Alternatively, \psr\ could be
equally as young as 1E~2259$+$586 and may not possess a light-element
insulating layer, in spite of the ease with which Heyl \& Hernquist
(1997a) suggest them to be formed.

\section{Conclusions}

In summary, we have set an upper limit on X-ray emission from the newly
observed high magnetic field radio pulsar \psr.  The upper limit on the
flux implies a luminosity in X-rays well below those of any known AXP.
This conclusion is independent of any beaming given the low pulsed
fraction of the known AXPs, and is robust to modest flux variability.
This argues that any magnetar mechanism invoked to explain X-ray
emission from AXPs must depend on more than observed $P$ and $\dot{P}$,
hence on more than merely inferred B field.  In the context of the two
particular magnetar hypotheses discussed above, \psr\ must be
considerably older than any of the known AXPs, including \axp, in spite
of characteristic ages that indicate otherwise.  This is consistent
with \axp\ being a much younger object than its characteristic age
suggests, in agreement with its association with CTB~109.
Additionally, it implies that its spin-down history has deviated
significantly from dipole braking, or that it had a large (few seconds)
initial spin period.  SGRs and AXPs may thus have markedly different
evolutionary paths from radio pulsars, including even the extreme
members of the latter population.  Thus, the proximity of SGRs and AXPs
to any radio pulsar in $P-\dot{P}$ phase space must be considered
merely coincidental and caution must be used when comparing the
different classes of neutron stars.

\bigskip

\acknowledgments 
We thank D. Helfand and an anonymous referee for helpful comments
on the manuscript.  We also thank B. Gaensler for useful discussions.
We have used the NASA-maintained HEASARC web site for archival
data retrieval and subsequent analysis.  M.~J.~P is supported in part
by NASA under Contract NASA-37716.  V.~M.~K. is supported in part by
the NASA Long Term Space Astrophysics program under grant NAGS$-$8063.


\bigskip 

\clearpage

\clearpage

\nocite{tmlc95} \nocite{ymj99} 

\clearpage

\begin{deluxetable}{l r r}
\small
\tablewidth{0pt}
\tablecaption{Comparison of \psr\ and \axp. \label{tab1}}
\tablehead{ \colhead{Parameter} & \colhead{\psr} & \colhead{\axp} }
\startdata
Spin period, $P$ (s)             & 4.0 & 7.0 \nl
Period derivative, $\dot{P}$ & $7.4 \times 10^{-13}$ & $4.9 \times 10^{-13}$ \nl
Surface Magnetic Field, $B$ (G) & $5.5 \times 10^{13}$ & $5.9 \times 10^{13}$ \nl
Characteristic age, $P/2\dot{P}$ (kyr)& 85 & 230 \nl
Spin-down Luminosity, $\dot{E}$ (\LX) & $4.7 \times 10^{32}$ & $5.7 \times 10^{31}$ \nl
Reference & Camilo et al. (2000) & Kaspi, Chakrabarty, \nl
& & \& Steinberger (1999) \nl
\enddata
\end{deluxetable}

\begin{deluxetable}{l r c   c r}
\small
\tablewidth{0pt}
\tablecaption{Pulse Properties of Known AXPs. \label{tab2}}
\tablehead{ \colhead{Pulsar} & \colhead{{\it f} \tablenotemark{1}} & \colhead{Passband\tablenotemark{2}} & \colhead{Shape} & \colhead{Ref\tablenotemark{3}} }
\startdata
4U~0142$+$61     & 7\%        & $0.5-1.5$ & two broad symmetric peaks    & 1,2 \nl
                 & 13\%       & $\phn\phd4-10\phd$    & single peak               & 1,2 \nl
1E~1048.1$-$5937 & 70\% & $0.5-1.5$ & single sinusoidal peak       & 3 \nl
                 & 70\% & $1.5-4.0$ & single sinusoidal peak  & 3 \nl
                 & 70\% & $4.0-8.0$ & single sinusoidal peak  & 3 \nl
		 & 70\% & $0.5-10\phd$ & single sinusoidal peak & 4 \nl
1E~1841$-$045    & 15\% & $\phn\phd1-10\phd$    & two broad overlapping peaks & 5,6  \nl
1E~2259$+$586    & 30\% & $0.1-2.4$ & two broad asymmetric peaks & 7 \nl
	         & 35\% & $\phn\phd1-10\phd$  & two broad asymmetric peaks & 8 \nl
1RXS~J170849.0$-$400910 & 38\% & $0.1-2.4$ & single sinusoidal peak & 9 \nl
	   & 50\% & $\phn\phd2-4\phd\phn$  & single sinusoidal peak & 10 \nl
	   & 50\% & $\phn\phd4-10\phd$     & single broad peak & 10 \nl
\tablenotetext{1}{See $\S$3.2 for the definition of {\it f}, the pulsed fraction. 
When not stated by the authors, we estimate {\it f} from the published 
pulse profiles.  (Note:  The previously reported value of 30\% for 
1E~1841$-$045 arises do a different definition of {\it f}.)}
\tablenotetext{2}{Units are keV.}
\tablenotetext{3}{References--(1) White et al. 1996, (2) Israel et al. 1999b, 
(3) Corbet \& Mihara 1997, (4) \cite{opmi98}, (5) Vasisht \& Gotthelf 1997, (6) Gotthelf, Vasisht 
\& Dotani 1999, (7) Rho \& Petre 1997, (8) Corbet et al. 1995, (9) Isreal et al.
1999a, (10) Sugizaki et al. 1997. 
}
\nocite{gvd99} \nocite{ics+99} \nocite{ioa+99}
\enddata
\end{deluxetable}

\end{document}